\documentclass[apj]{emulateapj}
\usepackage{apjfonts}

\begin{document}

\title{Probabilistic Cross-Identification of Astronomical Sources}
\journalinfo{To appear in the Astrophysical Journal}

\author{Tam\'as Budav\'ari and Alexander S. Szalay}
\affil{Dept.\ of Physics and Astronomy, The Johns Hopkins University, 3400 North Charles Street, Baltimore, MD 21218}
\affil{Max-Planck-Institute f\"ur Astrophysik, Karl-Schwarzschild-Strasse 1, 85748 Garching, Germany}

\shortauthors{Budav\'ari and Szalay}
\shorttitle{Probabilistic Cross-Identification of Astronomical Sources}

\begin{abstract}
We present a general probabilistic formalism for cross-identifying
astronomical point sources in multiple observations. Our Bayesian approach,
symmetric in all observations, is
the foundation of a unified framework for object matching, where not only
spatial information, but physical properties, such as
colors, redshift and luminosity,
can also be considered in a natural way.
We provide a practical recipe to implement an efficient recursive algorithm to
evaluate the Bayes factor over a set of catalogs with known circular errors
in positions.
This new methodology is crucial for
studies leveraging the synergy of today's multi-wavelength observations and
to enter the time-domain science of the upcoming survey telescopes.
\end{abstract}


\section{Motivation} \label{sec:intro}

Observational astronomy has changed drammatically over the last decade.
With the introduction of large-format, high-resolution detectors at
all wavelengths of the electromagnetic spectrum,
astronomers now face an avalanche of data pouring from the instruments of
dedicated telescopes. While most imaging surveys today obtain multicolor
information, no one telescope can cover the entire spectrum because the
physics of the detectors is very different at different frequencies.
To fully utilize the available observations, e.g.,
to boost the chances of discovering new kinds of sources, and understanding
the underlying physical relations of object properties in a statistical way,
one needs to merge the datasets of various telescopes by federating the archives.
The Virtual Observatory
initiative spearheaded by the International Virtual Observatory Alliance%
\footnote{http://www.ivoa.net} (IVOA)
is pursuing automated data exchange protocols with catalog cross-identification,
and the US National Virtual Observatory%
\footnote{http://us-vo.org} (NVO) is building tools, e.g., Open SkyQuery
\citep{budavari04} to facilitate a standard unified framework.
The key step in
the process is the cross-identification of the sources in multiple catalogs
to link observations of one telescope to other's.
Previous attempts to alleviate the problem utilized likelihood analysis
\citep{ss92} and machine learning (ML) techniques \citep{rohde06} that
addressed specific issues of the matching problem of two catalogs.
\citet{mann97} successfully applied the former likelihood ratio method
to associate sources in the
Infrared Space Observatory and the Hubble Deep Field catalogs,
and the ML techniques were used to study the SuperCOSMOS observations
and HI Parkes All Sky Survey.

Today astronomers typically join two catalogs by setting some threshold on the angular
separation of sources that is motivated by the astrometric accuracies of the
datasets involved. When more than two catalogs are to be crossmatched,
astronomers often hatch a chaining rule based on the implicit
prior knowledge about the sources. For example, one might decide to match
all lower-accuracy datasets to the best one, or to go from wavelength to
wavelength, hoping that the sources do not change significantly over a shorter
wavelength range.
The problem with these traditional ways is not that they are based on implicit
assumptions and intuitions but that they are not symmetric. While the pairwise
matches might be acceptable, there is no guarantee, or any measure of quality,
that the elected final matches are plausible or if the list is complete.
After all picking a different order of pairwise matching would yield a different
catalog.

We need algorithms that are symmetric in the catalogs and provide a reliable
measure of quality that one can use to exclude or downweight unlikely combinations of
sources. We need a unified framework, where on top of the spatial
information, other measurements can also be incorporated along with explicit
models and physical priors.
In Section~\ref{sec:bf} we discuss the Bayesian approach to address these issues,
and in Section~\ref{sec:norm} the spherical normal distribution is studied.
In Sections~\ref{sec:physics} we demonstrate how to
include other observational evidence such as from multicolor photometric measurements.
Section~\ref{sec:thres} focuses on the effects of a limited field of view on
the observational evidence, the prior and posterior probabilities.
In Section~\ref{sec:prac} an efficient implementation of the framework is
described in the detail, and Section~\ref{sec:sum} concludes our study.

Throughout the paper we follow the usual convention of using the lower-case $p$ symbol
for representing probability density functions and the capital $P$ symbol for probabilities.

\section{Observational Evidence} \label{sec:bf}

Often Bayesian analysis is refered to as the calculus of belief, however, it
should rather be thought of as the calculus of observational evidence.
When presented with a series of observed positions, one would like to know
whether they are truely from the same source. If the coordinates are scattered
all over the celestial sphere, it seems very unlikely that they are
measurements of the same astronomical object, but when the coordinates are only
a tiny fraction of an arcsecond apart, we ``know'' that we found a good match.
How good is that match? Or what is the evidence that it is a match?

\subsection{Modelling the Astrometry}

First let us examine what astrometric precision means. In the process of
calibrating the positions in a catalog of extracted sources, one can
characterize the properties of the observations by comparing the positions to
astrometric standards, and even correct for systematic offsets. Yet, there
remains a random scatter around the true positions. This uncertainty is often
modelled as a normal distribution, and catalogs would quote a
single $\sigma$-value for their accuracy, e.g., $\sigma=0.1$ arcseconds.
In general, our understanding of the astrometry is described by a
probability density function (PDF) that may even vary on the sky.
We parameterize our model $M$ that the object is on the celestial sphere using a
three-dimensional normal vector $\vec{m}$, and write
$p(\vec{x}|\vec{m},M)$ for the probability density that an object at its
true location $\vec{m}$ is observed at a position $\vec{x}$.
As any PDF, this function is normalized,
\begin{equation}
\int\,p(\vec{x}|\vec{m},M)\,d^3\!x = 1
\end{equation}
Now we take a single source observed at $\vec{x}_1$ and apply Bayes'
theorem to find the posterior density of the true location $\vec{m}$ given
the obtained data,
\begin{equation}
p(\vec{m}|\vec{x}_1,M)=\frac{p(\vec{x}_1|\vec{m},M)p(\vec{m}|M)}{p(\vec{x}_1|M)}
\end{equation}
where the trivial prior $p(\vec{m}|M)$ of $\vec{m}$ being on the
celestial sphere is expressed with Dirac's $\delta$-symbol,
\begin{equation}
p(\vec{m}|M) = \frac{1}{4\pi} \delta(|\vec{m}|-1)
\end{equation}
and the normalizing constant guarantees the law of total probability,
\begin{equation}
p(\vec{x}_1|M) = \int p(\vec{m}|M)\,p(\vec{x}_1|\vec{m},M)\,d^3\!m
\end{equation}
Another interesting direct application is the calculation of the chance that
we find a visible object at position $\vec{m}$ in
a given footprint. If the angular window function is $\Omega$, this probability
is simply
\begin{equation}
P(\Omega|\vec{m},M) = \int_\Omega p(\vec{x}|\vec{m},M)\,d^3\!x
\end{equation}
which one can use to infer the PDF on the true position by applying
Bayes' rule
\begin{equation} \label{eq:win}
p(\vec{m}|M_{\Omega}) \equiv{}\, p(\vec{m}|\Omega,M) = \frac{p(\vec{m}|M)P(\Omega|\vec{m},M)}
{\displaystyle\int p(\vec{m}|M)P(\Omega|\vec{m},M)\,d^3\!m}
\end{equation}
This is our best understanding of where an object might be on the sky
(prior to measuring its actual position)
that is
seen in the specified $\Omega$ footprint assuming astronometric precision
$p(\vec{x}|\vec{m},M)$ derived from the calibration.

\subsection{The Bayes Factor}

With multiple observations through various instruments of possibly different
astrometric accuracies, we now turn to compute the evidence that all
observations are from the same source.
We introduce the Bayes factor to test this hypothesis $H$ against the case
when separate sources are possible, $K$.
After the observation are obtained,
\mbox{$D=\{\vec{x}_1,\vec{x}_2, \dots, \vec{x}_n\}$} locations on the sky,
we compute the ratio of the posterior and prior probabilities of each hypothesis.
The Bayes factor is defined as the ratio of these odds,
\begin{eqnarray} \label{eq:bf1}
B(H,K|D)&=&\left(\frac{P(H|D)}{P(H)}\right)\Big/\left(\frac{P(K|D)}{P(K)}\right)
\end{eqnarray}
which, after applying Bayes' theorem, becomes
\begin{eqnarray} \label{eq:bf2}
B(H,K|D) & = & \frac{p(D|H)}{p(D|K)}
\end{eqnarray}
for continuous observables.
The actual calculation is done by parameterizing the two models $H$ and $K$, and
integrating the likelihood functions for the entire configuration space.
Our hypothesis $H$ says that the positions are from a single source, thus
can be parameterized by a single common location $\vec{m}$.
Due to the independence of the measurements in $D$, the joint PDF is just
the product of the astrometric precisions $p_1, p_2, \dots, p_n$,
and the integral simplifies to
\begin{equation}
p(D|H) = \int p(\vec{m}|H) \prod_{i=1}^{n} p_i(\vec{x}_i|\vec{m},H)\,d^3\!m
\end{equation}
On the other hand, the alternative hypothesis $K$ is parameterized by
separate $\{\vec{m}_i\}$
positions, and the integral factorizes into the product of the independent
components
\begin{equation}
p(D|K)  = \prod_{i=1}^{n} \left\{\int p(\vec{m}_i|K)\,p_i(\vec{x}_i|\vec{m}_i,K)\,d^3\!m_i \right\}
\end{equation}
When the Bayes factor is large, the observations support the hypothesis that the
association is a match, if it is in the order of unity, the evidence is not convincing,
and finally if the ratio is less than one, the data prefers the alternative
hypothesis.

\section{The Normal Distribution} \label{sec:norm}

Normal distributions emerge often in nature, where a number of effects play
roles in shaping up the probability density, cf.~the Central Limit theorem.
Although many of the usual arguments do not hold over closed topological
manifolds,
e.g., the Central Limit theorem leads to isotropic distribution on the circle \citep{levy39},
it is possible to introduce an analogue to the normal distribution
function on the sphere \citep{fisher53,breitenberger63}.
The spherical normal distribution is often elected to
characterize the precision of astronomy observations, hence it is of great
importance to understand its properties, and to apply the Bayesian framework
described in the previous section.

The spherical normal distribution in its normalized form using the previous
3-D vector notation is written as
\begin{equation}
N(\vec{x}|\vec{m},w) = \frac{w\,\delta(|\vec{x}|\!-\!1)}{4\pi \sinh w}\
           \exp \left( {w\,\vec{m}\vec{x}} \right)
\end{equation}
where the weight $w$ is typically very large. When this is the case, the weight
is related to
the more intuitive precision parameter $\sigma$ by the equation
\begin{equation}
w = 1/\sigma^2
\end{equation}
For example, when $\sigma$ is in the order of an arcsecond, the weight takes values of
$\sim 10^{10}$.
Having observed a set of positions independently with corresponding weights,
we can compute
the Bayes factor for the two hypotheses $H$ and $K$ introduced earlier.
Because the function $N(\vec{x}|\vec{m}, w) p(\vec{m}|M)$ is symmetric in
$\vec{x}$ and $\vec{m}$ for the trivial prior, and the PDFs are normalized,
the Bayes factor is computed analytically, and becomes
\begin{equation} \label{eq:sinh}
B(H,K|D)=\frac{\sinh{}w}{w}\,\prod_{i=1}^{n} \frac{w_i}{\sinh{}w_i}
\end{equation}
with
\begin{equation}
w = \left| \sum_{i=1}^{n} w_i \vec{x}_i \right|
\end{equation}
where we exploit the fact that the product of normal distributions has the same
functional form. A detailed derivation is given in Appendix~\ref{app:1}.

\begin{figure}[bh]
\epsscale{1}
\plotone{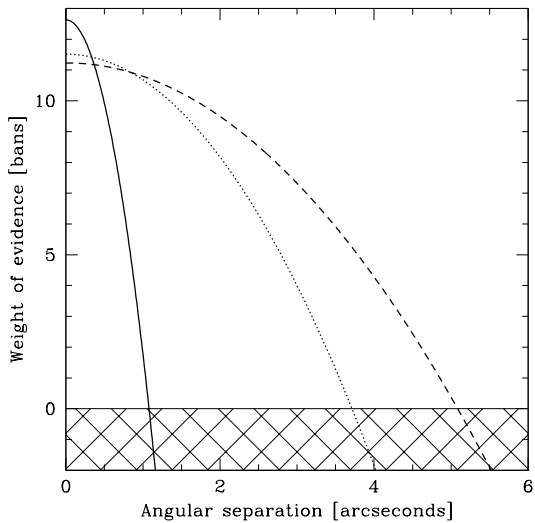}
\caption{The weight of evidence as a function of angular separation
for the three cases of matching two catalogs of $\sigma_1=0.1"$ and $\sigma_2=0.5"$
to each other and to themselves. For example, matching SDSS and GALEX sources:
SDSS--SDSS (solid), SDSS--GALEX (dotted) and GALEX--GALEX (dashed).}
\label{fig:ex}
\end{figure}

In case of only two observations, this weight depends on the astrometric
precisions and the angle $\psi$ between the positions
\begin{equation}
w = \sqrt{w_1^2 + w_2^2 + 2 w_1 w_2 \cos \psi}
\end{equation}
For the typical large weights and small angular separations between the
measurements, we get
\begin{equation}
B = \frac{2}{\sigma^2_1 + \sigma^2_2}
         \exp \left\{-\frac{\psi^2}{2(\sigma^2_1 + \sigma^2_2)} \right\}
\end{equation}
In Figure~\ref{fig:ex} the 10-based logarithm of the Bayes factor, also known as
the weight of evidence, is shown as a function of angular separation
for the three cases of matching two catalogs of $\sigma_1=0.1"$ and $\sigma_2=0.5"$
to each other and to themselves. This is the problem of matching the Sloan
Digital Sky Survey \citep[SDSS;][]{york,pier} and the Galaxy Evolution Explorer
\citep[GALEX;][]{martin,morrissey} science archives.

\begin{deluxetable}{cccr|cccr}
\tablecolumns{8}
\small
\tablewidth{0pt}
\tablecaption{Weight of evidence for three surveys
as a function of the angular separations in $\sigma=\sigma_1=\sigma_2=\sigma_3=0.1"$ units}
\tablehead{
\colhead{$\psi_{12}$} &
\colhead{$\psi_{23}$} &
\colhead{$\psi_{31}$} &
\colhead{$W$} &
\colhead{$\psi_{12}$} &
\colhead{$\psi_{23}$} &
\colhead{$\psi_{31}$} &
\colhead{$W$}
}
\startdata
 0 &  0 &  0 &  25.38  &   0 &  0 &  0 &  25.38 \\
 1 &  1 &  1 &  25.17  &   0 &  1 &  1 &  25.24 \\
 2 &  2 &  2 &  24.51  &   0 &  2 &  2 &  24.80 \\
 3 &  3 &  3 &  23.43  &   0 &  3 &  3 &  24.08 \\
 4 &  4 &  4 &  21.91  &   0 &  4 &  4 &  23.07 \\
 5 &  5 &  5 &  19.95  &   0 &  5 &  5 &  21.76 \\
 6 &  6 &  6 &  17.57  &   0 &  6 &  6 &  20.17 \\
 7 &  7 &  7 &  14.74  &   0 &  7 &  7 &  18.29 \\
 8 &  8 &  8 &  11.49  &   0 &  8 &  8 &  16.12 \\
 9 &  9 &  9 &   7.79  &   0 &  9 &  9 &  13.66 \\
10 & 10 & 10 &   3.67  &   0 & 10 & 10 &  10.91 \\
11 & 11 & 11 &  -0.89  &   0 & 11 & 11 &   7.87 \\
12 & 12 & 12 &  -5.89  &   0 & 12 & 12 &   4.54 \\
13 & 13 & 13 & -11.32  &   0 & 13 & 13 &   0.92 \\
14 & 14 & 14 & -17.18  &   0 & 14 & 14 &  -2.99 \\
\enddata
\label{tbl:b3}
\end{deluxetable}

Matching three catalogs also makes an interesting case study for the various
potential configurations of three positions. The Bayes factor for this case,
in the same limit as previously, takes the form of
\begin{equation}
B = \frac {4 \exp \left\{ -\frac
              { \sigma^2_3 \psi_{12}^2 + \sigma^2_1 \psi_{23}^2 + \sigma^2_2 \psi_{31}^2 }
              { 2(\sigma^2_1 \sigma^2_2 + \sigma^2_2 \sigma^2_3 + \sigma^2_3 \sigma^2_1)} \right\}}
           {\sigma^2_1 \sigma^2_2 + \sigma^2_2 \sigma^2_3 + \sigma^2_3 \sigma^2_1}
\end{equation}
In Table~\ref{tbl:b3} the weight of evidence is shown for various configurations
from the matching of three similar catalogs with equal astrometric accuracies,
$\sigma_1 = \sigma_2 = \sigma_3 = 0.1"$ (separations listed in $\sigma$ units.)
The astrometric precision was chosen to match the nominal SDSS limitations.

In general, the Bayes factor for the typical large weights and small
angular separations takes the form of
\begin{equation}
\label{eq:gensml}
B = 2^{n-1} \frac{\prod w_i}{\sum w_i} \exp \left\{ -
  \frac{\sum_{i<j} w_i w_j \psi_{ij}^2}{2\sum{w_i}} \right\}
\end{equation}
where all summations and products run on the members of the tuple from
the $n$ number of catalogs; see Appendix~\ref{app:2} for the details of the
calculation.

In scenarios where individual errors are different or even anisotropic,
one can generalize our expression in a fairly straightforward manner
in the above approximation. Instead of the scalar weight, one can use
the inverse of the covariance matrix, however, the elegant simplicity
of the expressions is sacrificed.

\section{Folding in the Physics}  \label{sec:physics}

Naturally the formalism introduced in Section~\ref{sec:bf} is not specific
to astrometric observations.
In fact, it is rather straightforward to fold other measured quantities
into the calculations.
This is especially important when dealing with multiple matches.
Picking the ``correct'' combination of sources from various spatially similar
configurations is a degenerate problem
that requires extra information to resolve. The use of photometric
information is a natural choice for its wide availability,
however, its application requires further assumptions on the spectral energy
distributions (SEDs). Often models exist to help out with the solution,
but extra caution is needed to avoid any undesirable effect. For example,
when the goal is to discover new types of objects with unknown SEDs, one
should not apply known SEDs as priors but rather look for combinations that
are likely matches based on spatial detections but excluded by SED modelling.

As a demonstration of these ideas, let us apply the introduced Bayesian framework to photometric
measurements in various passbands.
The ingredients include the following further explicit models:
\begin{itemize}
\item[1] Model $S$ for the spectrum energy distributions,
e.g., by \citet{bc03}, described by a set of parameters, $\vec{\eta}$:
$s(\lambda|\vec{\eta},S)$ along with the corresponding $p(\vec{\eta}|S)$ priors;
\item[2] Model $R$ for the transmission of the passbands to calculate
simulated fluxes $\vec{\gamma}(\vec{\eta}|S,R)$
by integrating the SEDs $s(\lambda|\vec{\eta},S)$ with the
appropriate response functions; and
\item[3] Model $C$ for the uncertainty of the catalog
from the photometric calibration, $p(\vec{g}|\vec{\gamma}, C)$, where $\vec{g}$
is the observed flux set and $\vec{\gamma}$ is the true.
\end{itemize}
These separate models can be folded into a single model $M$, for simplicity,
so one can write $p(\vec{g}|\vec{\eta},M)$ for the probability density of measuring
$\vec{g}$ fluxes for an object with $\vec{\eta}$ physical properties of $S$
seen through the filters in $R$ with the $C$ photometric accuracy.
%
The Bayes factor for the photometry in the face of the observed fluxes
$D'=\{\vec{g}_1,\vec{g}_2,\dots{},\vec{g}_n\}$, similarly to the astrometric
formulas, is given by the ratio
\begin{equation}
B(H,K|D') = \frac{\displaystyle%
\int p(\vec{\eta}|H) \prod_{i=1}^{n} p_i(\vec{g}_i|\vec{\eta},H)\,d^r\!\eta
 }{\displaystyle%
\prod_{i=1}^{n} \left\{\int  p(\vec{\eta}_i|K)\,p_i(\vec{g_i}|\vec{\eta}_i,K)\,d^r\!\eta_i\right\}
}
\end{equation}

In the simplest case, $S$ is parameterized by a discrete spectral type $T$,
the redshift $z$ and an overall scaling factor for the brightness, $\alpha$:
\begin{equation}
\vec{\gamma} = \alpha \vec{f}(T,z)
\end{equation}
where $\vec{f}$ is a vector of the simulated photometry in the various
passbands.
Photometric uncertainties are often assumed to be Gaussian with a diagonal
covariance matrix of elements $\sigma_l^2$, where $l$ runs on the
$L$ number of passbands. After substitution, we arrive at the familiar formula of
\begin{equation}
p(\vec{g}|\vec{\eta},M) = \frac{1}{\cal{}N}
  \exp \left\{-\sum_{l=1}^{L} \frac{[g_l-\alpha f_l(T,z)]^2}{2\sigma_l^2} \right\}
\end{equation}
where constant $\cal{}N$ is the usual normalization factor of the multivariate
normal distribution, which in our special case is just
${\cal{}N} = (2\pi)^{L/2} \sigma_1 \sigma_2 \cdots \sigma_L$.
Integrating these models to get the Bayes factor is a very similar problem
to template fitting photometric redshift estimation. In fact, the two
procedures can be done in a self-consistent way within the same application.
Naturally, spectroscopic redshift measurements can be directly incorporated
in this analysis, when available, but other data can also enter in
a straightforward manner.

The Bayesian analysis is inherently recursive. As soon as we obtain new
measurements, and compute the posterior probability, that becomes the
prior for subsequent studies. This is an extremely powerful property, and
simplifies the computations enormously.
A consequence of this is that the combined Bayes factor of the astrometric and
photometric measurements is simply the product of the two,
\begin{equation} \label{eq:bb}
B = B_{\rm{}pos} \cdot B_{\rm{}phot}
\end{equation}
as also seen from the Bayes factor's definition.
This means that one can just do the spatial join first, and consider
additional measurements and physical priors in subsequent steps, if needed.

\section{From Priors to Posteriors} \label{sec:thres}

The Bayes factor naturally relates the prior and posterior probabilities. When
$K$ is the complementary hypothesis of $H$,
the posterior probability is
\begin{equation} \label{eq:post}
P(H|D) = \left[ 1 + \frac{1-P(H)}{B\,P(H)} \right]^{-1}
\end{equation}
which, in the limit of vanishing priors, becomes
\begin{equation} \label{eq:post2}
P(H|D) = \frac{B\,P(H)}{1+B\,P(H)}
\end{equation}
To make a definitive decision on whether a set of detections
should be considered a match, one would like to set a limit on the posterior
probability and derive the Bayes factor threshold from that, however,
this can only be done with an initial estimate of the prior.

\subsection{The Prior and the Selection Function}

The prior probability depends on the angular and radial selection functions
of the observations.
If the visible universe contains $N$ objects,
and we select two of them at random, the probability of picking the
same object is $1/N$. When selecting $n$ objects, the probability
is $1/N^{n-1}$. A limited field of view shrinks the observable volume,
hence decreases the number of objects, and increases the prior probability.
When the angular selection functions of the catalogs overlap only partially
then one can just consider the intersection of the sky coverage
and the smaller number of sources within.

The various radial
selection functions also have a significant role, and make the situation
more complicated. In order to consider their effect, one has to estimate
the overlap of the selections in the input catalogs. Every
catalog has observational constraints, other than the field of view,
like flux limits,
that set the radial selection function. The superset
of these contraints defines the restrictions on the {\em{}overlap catalog}.
Let $N_\star$
denote the number of objects in that catalog.
In this general case, the prior probability takes the form of
\begin{equation} \label{eq:uniform}
P(H) = N_\star \Big/ \prod_i^n N_i
\end{equation}
When the limitations are identical, all catalogs have equal number of
objects, $N_\star = N_1 = \dots{} = N_n$, and we get back the same formula
of $P(H)=1/N^{n-1}$ as before, but when, for example, one catalog consists of only
low-redshift galaxies (e.g., $z<0.2$), and the other has high-redshift quasars 
(e.g., $z>3$), there is no overlap between the two radial selection functions,
hence $N_\star=0$, which means $P(H)=0$. One can get vanishing priors even if
the redshift histograms overlap significantly, e.g., two catalogs of
red (e.g., $u-g<2$) and blue galaxies ($u-g>2$).
In general, all these complex selection effects
are captured in a single scalar quantity, $N_\star$, which is estimated based on
prior physical knowledge, e.g., by using template SEDs
and the known characteristics of the input catalogs (e.g., the luminosity
functions), or
alternatively, when no prior information is available, one can invoke
self-consistency arguments to derive it; see later.
We now rewrite the prior with the surface densities, $\nu = N/\Omega$,
or the scaled number of objects for the entire sky, $\rho=4\pi\nu$, as
\begin{equation} \label{eq:omega}
P(H) = \frac{\nu_\star}{\prod \nu_i}\,\Omega^{1-n}
 = \frac{\rho_\star}{\prod \rho_i} \left(\frac{\Omega}{4\pi}\right)^{1-n}
\end{equation}
This formula also provides a straightforward way to include a model
for varying surface density on the sky, e.g., for stars,
where $\nu = \nu(\vec{x})$. In this case, a constant limiting posterior
probability yields a varying threshold on the Bayes factor as a function of
the position on the celestial sphere.

\subsection{The Bayes Factor and the Window Function}

The field of view not only changes the prior probabilities but also
modifies the Bayes factor.
When the window function is known, one can
refine the prior probability density that enters the integral of the
numerator and denominator of the Bayes factor. This is done by adopting
eq.~\ref{eq:win} as the prior.
In first order, for the typical catalogs with large weights (high accuracy)
and large contiguous observation areas, this new prior is uniform over the window
function, neglecting the fuzzy boundary, except scaled by the area coverage
\begin{equation}
p(\vec{m}|M_\Omega) = \frac{\Omega(\vec{m})}{\Omega}\,\delta(|\vec{m}|-1)
\end{equation}
where again $\Omega(\vec{m})$ is the window function that takes the value 1
when $\vec{m}$ is inside and 0 otherwise, and $\Omega$ is its area.
The Bayes factor inside the footprint will be essentially same as before in
eq.~\ref{eq:sinh} but also scaled with these fractional coverage:
\begin{equation} \label{eq:wsinh}
B = \left(\frac{\Omega}{4\pi}\right)^{n-1}
  \frac{\sinh{}w}{w}\,\prod_{i=1}^{n} \frac{w_i}{\sinh{}w_i}
\end{equation}
The edge effect modifies this only for a tiny fraction of the objects at
the boundary of the observations. The proper integral is, of course, much
more expensive than this analytical formula, but can be evaluated or
re-evaluated, e.g., by an MCMC algorithm.

For the typical small priors, the posterior depends only on the product of the
Bayes factor and the prior; see eq.~\ref{eq:post2}. This means that the footprint
effect cancels out in the posterior
probability; cf.\ eqs.~\ref{eq:omega} and \ref{eq:wsinh}.
Hence it is still sensible to just simply use the all-sky formula
in eq.~\ref{eq:sinh} and \ref{eq:gensml} as
long as the prior is written accordingly, i.e., $\rho_\star/\rho_1\rho_2\cdots\rho_n$.
From the data providers' point of view, who often do not know the field
of view, e.g., the legacy catalogs in VizieR \citep{vizier}
or the NASA/IPAC Extragalactic Database \citep[NED;][]{ned},
the best quantity to
publish along with the matched tuples is also the analytic all-sky Bayes factor,
so researchers can incorporate their own prior knowledge, and set the thresholds
on the posterior accordingly that are often specific to the science application.

\subsection{Self-Consistent Estimation}

In principle, the cross-identification process is now complete,
one just has to formulate the prior, possibly varying on the sky,
and set a threshold on the posterior probabilities to select the matches.
For the ignorant without a priori knowledge, these are not completely independent
choices, and, at least in the limit when all observables are being considered
in the Bayes factors, could be derived from requirements of a self-consistent
field theory.
When prior knowledge is available and dictates a preference, one could
and probably should still check for the consistency outlined here to understand
the discrepancies, if any.

The formula for the prior in eq.~\ref{eq:uniform}
is in fact equivalent to stating that $P(H)$ is constant and
\begin{equation}
\sum P(H) = N_\star
\end{equation}
where the summation runs over the direct product of all
sets of sources in the $n$ catalogs, i.e., all possible combinations
of detections with $N_1 N_2\cdots{}N_n$ contributions.
The self-consistency argument requires that
\begin{equation} \label{eq:self}
\sum P(H|D) = N_\star
\end{equation}
which is an equation for $N_\star$ that can be solved by, e.g.,
some iterative approximation method starting from
an initial value of $N_\star = \min \left\{ N_i \right\}$.
Initial experiments support our expectations that these procedures indeed converge
very rapidly, only in a few iterations,
and are insensitive to the matching limit once the Bayes factor is less than unity.
For varying unknown priors one can use some sky tesselation schemes,
such as HEALPix \citep{healpix}, Igloo \citep{igloo} or HTM \citep{szalay05},
and estimate a piecewise constant prior (uniform in the cells) using
the same methodology. Naturally other more sophisticated models can also be
used in the same spirit, e.g., specific functional forms or smoothing to limit
the gradient, as well as tapered windows when required.

The threshold on the posterior, $P_T$, can also be established in a consistent way.
Here the requirement is that
\begin{equation}
\sum_{P(H|D)>P_T}\!\!\!\!\!\!\!1 = N_\star
\end{equation}
This is equivalent to applying a Bayes classifier.
By changing the right hand side of the above equation, it is possible to make
the selection more restrictive or less depending on the scientific goal.
In the case, where the prior changes on the sky and eq.~\ref{eq:self} is solved in cells of
some pixelization, one can still just use a single $P_T$ limit obtained from
the entire catalog by ensuring that the total number of objects are consistent.
The counts in individual cells may not be perfectly recovered but, if the prior
is right, there should be no significant trends.

\section{Practical Considerations}  \label{sec:prac}

The question remains how to evaluate the Bayes factor efficiently for multiple
catalogs without considering all possible combinations of sources.
Fast algorithms exist to match two sets of point sources
using an angular separation limit \citep{budavari03,malik03,gray04,gray06,szalay05,nieto07}.
Ideally one would like to leverage the power of these two-way crossmatch
engines in a recursive manner, and get rid of unlikely combinations with small
Bayes factors as early as possible.

Matching two catalogs is straightforward; any Bayes factor limit corresponds
to a single distance cut, and hence our existing tools are adequate.
To go from $n$ number of catalogs to $n+1$, we need to make this
process iterative, and prune the match list step-by-step.
We do this by computing the overall Bayes factor in every step assuming that
all other subsequent catalogs will contribute sources at the best possible
position.
This optimization problem may be expensive to solve in general, but can
be analytically calculated in special cases, and for the spherical normal
distribution the solution is evident: the center position of the mode is
the correct choice.

In fact, for the normal distribution one can do even better. In every
step, a new catalog is added to the current sub-matches.
Since the product of normal distributions is still of the same functional form,
one can compute the Bayes factor as a
function of angular separation from that position, derive the limiting radius,
and utilize a two-way crossmatch engine for joining the current $k$-tuples with
the new $(k+1)^{\rm{}th}$ catalog using that threshold.
For this we rewrite the logarithm of the Bayes factor in eq.~\ref{eq:gensml},
in the more convenient form of
\begin{eqnarray} \label{eq:bfinc}
\ln B = \ln N - \frac{1}{2}\sum_{i=2}^n \frac{a_{i-1}}{a_i} w_i \Delta_i^2
\end{eqnarray}
with the newly introduced variables
\begin{eqnarray}
N & = & 2^{n-1} \frac{\prod w_i}{\sum w_i} \\
a_k & = & \sum_{i=1}^k w_i \\
\vec{\Delta}_i & = & \vec{x}_i - \vec{c}_{i-1}
\end{eqnarray}
where $\vec{c}_{k}$ is the unit vector of the best position for the current
$k$-tuple of sub-match,
\begin{equation}
\vec{c}_{k} = {\displaystyle\sum_{i=1}^k w_i \vec{x}_i}
   \Big/         {\left| \displaystyle\sum_{i=1}^k w_i \vec{x}_i \right|}
\end{equation}
With these we compute the weight of evidence in a recursive manner.
The iteration starts by substituting $\vec{c}_1 = \vec{x}_1$. In the $k^{\rm{}th}$ step,
the maximum search radius $\rho_{k+1}$ is computed from eq.~\ref{eq:bfinc} to yield
the Bayes factor threshold $B_{0}$ by assuming optimal matches from the subsequent
catalogs with vanishing $\Delta_i^2$ contributions,
\begin{equation}
b_{k+1} \rho_{k+1}^2 = 2\ln\frac{N}{B_{0}} - \sum_{i=2}^k b_i \Delta_i^2\ \ \
{\rm{}with}\ \ \
b_{k}=\frac{w_k a_{k-1}}{a_k}
\end{equation}
We assign every source within that radius to each $k$-tuple sub-match, and go
to the next catalog.
In general, the search radius will be different for every tuple for
their different spatial configurations. When the two-way matching algorithm
requires a fixed radius, one can take the maximum value in linear time, use
that more generous search radius in the matching,
and filter the result set later, just before going to the next catalog.
From catalog to catalog we propagate only the quantities that are
necessary to calculate the weight of evidence. The recursion formulas are
given by the following expressions:
\begin{eqnarray}
a_{k} &=& a_{k-1} + w_{k} \\
q_{k} &=& q_{k-1} + \frac{a_{k-1}}{a_{k}} w_{k} \Delta_{k}^2 \\
\vec{c}_{k} &=& \left({\vec{c}_{k-1} + \frac{w_{k}}{a_{k}} \vec{\Delta}_{k}}\right)
            \Big/ {\left|\vec{c}_{k-1} + \frac{w_{k}}{a_{k}} \vec{\Delta}_{k} \right|}
\end{eqnarray}
This stepwise method for evaluating the weight of evidence not only
provides an accurate match list that meets all our requirements enumerated
in Section~\ref{sec:intro}, e.g., symmetry in the catalogs,
but also exhibits the performance of the
current state-of-the-art two-way crossmatching tools.

\section{Summary}  \label{sec:sum}

We presented a general probabilistic formalism for cross-identifying astronomical
point sources. The framework is based on Bayesian hypothesis testing to decide
whether a series of observations truly belong to a single astronomical object.
The expression we derived is completely general, symmetric in all observations, and 
accommodates any model of the astrometric precision.
We introduced the spherical normal distribution, and calculated the Bayes factor
for the generic $n$-way matching problem both in the general case and in the
typical limit of high precision and small angular separations.
The cases of 2- and 3-way matching were studied in detail.
We discussed an efficient evaluation strategy of the Bayes factor that leverages
the power of existing high-performance two-way matching tools in a recursive manner,
yet, it provides accurate measurements of the observational evidence that are independent
of the order of the catalogs considered.
While the normal distribution is the simplest to work with for its unique
properties, other specific PDFs can be handled in the same spirit.
Our technique provides a natural mechanism to include other observed properties.
We demonstrated how multicolor survey data, even at different wavelengths,
can be utilized in the matching process by invoking SED models.
Morphological classification or redshift measurements, when available,
will also increase the accuracy of the results.

The beauty of our approach to the cross-identification problem is
that it completely separates the dependence on each parameter,
while providing the opportunity to incorporate
them in a fairly straightforward way.
Including expert knowledge about the physics of the objects in the analysis is
easily achievable by adopting the right priors,
and when such information is not available, self-consistency
arguments can guide the process to a stable solution in a few iterations.
With the pre-computed Bayes factors in the matched catalogs,
astronomers can define custom thresholds to derive specialized crossmatch catalogs
based on their own explicit assumptions. For example, using a database of
the same set of associations,
researchers can optimize for completeness of the galaxy population,
or even search for unusually red objects.

\appendix

\section{The Bayes factor and the Spherical Normal Distribution}
\label{app:1}

In this appendix we discuss the mathematical calculation of the Bayes factor
in the common case, when a spherical normal distribution is assumed for modelling
the astrometric accuracy. In addition we also adopt an all-sky prior in this
derivation.

The Bayes factor is the ratio of the likelihoods, $p(D|H)$ and $p(D|K)$,
where again $D$ represents the observed positions, $\left\{\vec{x}_i\right\}$.
\begin{eqnarray}
B = \frac{p(D|H)}{p(D|K)}
\end{eqnarray}
We recall that hypothesis $H$ is parameterized by a single position, $\vec{m}$ unit vector,
and $K$ is parameterized by a set of $n$ position vectors, $\left\{\vec{m}_i\right\}$.
The basic equations to start from are
\begin{eqnarray}
p(D|H) & = & \int\!d^3\!m\ p(\vec{m}|H)\ p(D|\vec{m},H) \\
p(D|K) & = & \int\!d^3\!m_1\!\int\!d^3\!m_2\dots\!\int\!d^3\!m_n
       \ p(\vec{m}_1|K)p(\vec{m}_2|K)\dots{}p(\vec{m}_n|K)\ p(D|\left\{\vec{m}_i\right\},K)
\end{eqnarray}
where
\begin{eqnarray}
p(\vec{m}|M)  &=&  \frac{\delta(|\vec{m}|\!-\!1)}{4\pi}  \\
p(\left\{\vec{x}_i\right\}|\vec{m},H)  &=& \prod_i^n N(\vec{x}_i|\vec{m},w_i) =
    \prod_i^n \frac{w_i\delta(|\vec{x}_i|\!-\!1)}{4\pi\sinh{w_i}}
        \exp\left({w_i\vec{x}_i\vec{m}}\right) \\
p(\left\{\vec{x}_i\right\}|\left\{\vec{m}_i\right\},K)  &=& \prod_i^n N(\vec{x}_i|\vec{m}_i,w_i) =
    \prod_i^n \frac{w_i\delta(|\vec{x}_i|\!-\!1)}{4\pi\sinh{w_i}}
        \exp\left({w_i\vec{x}_i\vec{m}_i}\right)
\end{eqnarray}
First we focus on hypothesis $H$
\begin{eqnarray}
p(D|H) & = & \int  d^3\!m \frac{\delta(|\vec{m}|\!-\!1)}{4\pi}
       \prod_i^n \frac{w_i\delta(|\vec{x}_i|\!-\!1)}{4\pi\sinh{w_i}} \exp\left({w_i\vec{x}_i\vec{m}}\right) \\
       & = & \left( \prod_i^n \frac{w_i\delta(|\vec{x}_i|-1)}{4\pi\sinh{w_i}} \right)
       \int  d^3\!m \frac{\delta(|\vec{m}|\!-\!1)}{4\pi} \exp\left({\sum_i^n w_i\vec{x}_i\vec{m}}\right)
\end{eqnarray}
introduce
\begin{equation}
w\vec{x} = \sum_i^n w_i\vec{x}_i
\end{equation}
where $\vec{x}$ is a unit vector, and write
\begin{eqnarray}
p(D|H) & = & \left( \prod_i^n \frac{w_i\delta(|\vec{x}_i|\!-\!1)}{4\pi\sinh{w_i}} \right)
       \int  d^3\!m \frac{\delta(|\vec{m}|\!-\!1)}{4\pi} \exp\left({w\vec{x}\vec{m}}\right) \\
       & = & \left( \frac{\sinh{w}}{w} \prod_i^n \frac{w_i\delta(|\vec{x}_i|\!-\!1)}{4\pi\sinh{w_i}} \right)
       \int  d^3\!m \frac{w\delta(|\vec{m}|\!-\!1)}{4\pi\sinh{w}} \exp\left({w\vec{x}\vec{m}}\right)
       \\
 & = & \frac{\sinh{w}}{w} \prod_i^n \frac{w_i}{\sinh{w_i}}\, \frac{\delta(|\vec{x}_i|\!-\!1)}{4\pi}
\end{eqnarray}
The likelihood of the alternative hypothesis $K$ is calculated similarly
\begin{eqnarray}
p(D|K) & = & \prod_i^n \int d^3\!m_i \frac{\delta(|\vec{m}_i|\!-\!1)}{4\pi}
       \frac{w_i\delta(|\vec{x}_i|\!-\!1)}{4\pi\sinh{w_i}}  \exp\left({w_i\vec{x}_i\vec{m_i}}\right) \\
       & = & \prod_i^n \frac{\delta(|\vec{x}_i|\!-\!1)}{4\pi}
\end{eqnarray}
Hence the Bayes factor is
\begin{equation}
B = \frac{\sinh{w}}{w} \prod_i^n \frac{w_i}{\sinh{w_i}}
\end{equation}
as also shown in eq.~\ref{eq:sinh}.

\section{High Astrometric Accuracy and Small Separations}
\label{app:2}

The astrometric precision of the actual observations is almost always extremely
high in the absolute sense,
so it is worth examining the approximation of the Bayes factor in this limit.
We also assume small angular separations. In the chain of equations below we only
use the ``$\approx$'' sign to signal new approximations.
We start from the previous result
\begin{eqnarray}
B & = & \frac{\sinh{w}}{w} \prod_i^n \frac{w_i}{\sinh{w_i}} \\
  & \approx & 2^{n-1} \frac{e^w}{w} \prod_i^n \frac{w_i}{e^{w_i}} \\
  & = & 2^{n-1} \frac{\prod w_i}{w}\,e^{w-\sum w_i} \\
  & = & 2^{n-1} \frac{\prod w_i}{w}\,e^{\sum w_i \left(\frac{w}{\sum w_i}-1\right)}
  \label{eq:bfsmlapp}
\end{eqnarray}
where we exploit the fact that all $w$ weights are large, hence the $\sinh{w}$ is
approximately $\frac{1}{2}\exp w$.
We proceed by calculating
\begin{eqnarray}
\left(\frac{w}{\sum_i w_i}\right)^2
& = & \frac{w^2}{\left(\sum_i w_i\right)^2} \\
& = & \frac{\sum_i w_i^2 + 2\sum_{i<j} w_i w_j \vec{x}_i\vec{x}_j}{\sum_i w_i^2 + 2\sum_{i<j} w_i w_j} \\
& = & \frac{\sum_i w_i^2 + 2\sum_{i<j} w_i w_j \cos\psi_{ij}}{\sum_i w_i^2 + 2\sum_{i<j} w_i w_j} \\
& \approx & \frac{\sum_i w_i^2 + 2\sum_{i<j} w_i w_j \left(1-\psi_{ij}^2/2\right)}{\sum_i w_i^2 + 2\sum_{i<j} w_i w_j} \\
& = & \frac{\sum_i w_i^2 + 2\sum_{i<j} w_i w_j -\sum_{i<j} w_i w_j \psi_{ij}^2}{\sum_i w_i^2 + 2\sum_{i<j} w_i w_j} \\
& = & 1-\frac{\sum_{i<j} w_i w_j \psi_{ij}^2}{\left(\sum_i w_i\right)^2}
\end{eqnarray}
After taking the square root of the above equation, we get
\begin{eqnarray}
\frac{w}{\sum_i w_i}
& \approx & 1 - \frac{\sum_{i<j} w_i w_j \psi_{ij}^2}{2\left(\sum_i w_i\right)^2}
\\
& \rm{and} & \nonumber \\
\sum_i w_i \left(\frac{w}{\sum_i w_i}-1\right)
& = & -\frac{\sum_{i<j} w_i w_j \psi_{ij}^2}{2\sum_i w_i}
\end{eqnarray}
From the above equations we also see that
\begin{equation}
\frac{1}{w} \approx \frac{1}{\sum_i w_i} \left(1+\frac{\sum_{i<j} w_i w_j \psi_{ij}^2}{2\left(\sum_i w_i\right)^2}\right)
\approx  \frac{1}{\sum_i w_i}
\end{equation}
in this context to only keep the leading term.
By substituting the above two equations to eq.~\ref{eq:bfsmlapp}, we arrive at our generic small angle result shown in
eq.~\ref{eq:gensml}.
The 2- and 3-way matching cases are straightforward specializations of the generic equation, where one substitutes
$w_i=1/\sigma_i^2$ to work out the simplified formulae.

\acknowledgements %
The authors are grateful for invaluable discussions
with Mar\'{\i}a Nieto-Santisteban, Istv\'an Csabai and Zolt\'an R\'acz
on various aspects of the topic,
and gladly acknowledge the generous support from the following
organizations:
Gordon and Betty Moore Foundation GBMF 554,
W. M. Keck Foundation KECK D322197,
NSF NVO AST-0122449,
NASA AISRP NNG05GB01G,
NASA GALEX 44G1071483,
Hungarian National Scientific Foundation OTKA-T047244,
European Research Training Network MRTN-CT-2004-503929.
Part of this research was done while A.~S. was a recipient of
an Alexander von Humboldt Fellowship at the MPA.

\end{document}